\documentclass{coresa} 

\usepackage[latin1]{inputenc} 
\usepackage[frenchb]{babel} 
\usepackage{t1enc,dfadobe}
\usepackage{cite}
\usepackage{caption}
\usepackage{subcaption}
\usepackage{color}
\usepackage{url}
\usepackage{flushend}

\usepackage{color}
\usepackage[colorlinks]{hyperref}
\hypersetup{
    colorlinks=true,                          
    linkcolor=blue, 
    citecolor=red, 
    urlcolor=blue  } 

\ifpdf \usepackage[pdftex]{graphicx} \pdfcompresslevel=9
\else \usepackage[dvips]{graphicx} \fi

\title[Preprint]{Design, Implementation and Simulation of a Cloud Computing System for Enhancing Real-time Video Services by using VANET and Onboard Navigation Systems}

			\author[K. Hammoudi	N. Ajam M. Kasraoui F. Dornaika K. Radhakrishnan K. Bandi Q. Cai S. Liu]%
       {K. Hammoudi$^{1,2}$ \hspace{0.3cm}	N. Ajam$^{1,2}$ \hspace{0.3cm} M. Kasraoui$^{1}$ \hspace{0.3cm} F. Dornaika$^{3,4}$ \hspace{0.3cm} K. Radhakrishnan$^{2,*}$ \hspace{0.3cm} K. Bandi$^{2,*}$ \hspace{0.3cm} Q. Cai$^{2,*}$ \hspace{0.3cm} S. Liu$^{2,*}$
        \\
         $^1$Research Institute on Embedded Electronic Systems (IRSEEM), IIS Group, Technop\^ole du Madrillet, St-Etienne-du-Rouvray, France \\
         $^2$ESIGELEC School of Engineering, Department of ICT ($^{*}$MS Students), Technop\^ole du Madrillet, St-Etienne-du-Rouvray, France\\
				 $^3$ Department of Computer Science and Artificial Intelligence, University of the Basque Country, San Sebastián, Spain \\
				 $^4$ IKERBASQUE, Basque Foundation for Science, Bilbao, Spain 
       }
			
\begin{document}

\maketitle

\begin{abstract}
 {\small Dans cet article, nous proposons une architecture pour le développement de systèmes de cloud computing nouveaux et expérimentaux. Le système proposé vise à renforcer les capacités de calcul, de communication et d'analyse de services de navigation routière par la fusion de plusieurs technologies indépendantes, à savoir les systèmes de navigation embarqués basés sur la vision, les systèmes de cloud computing de premier plan et les réseaux Ad-Hoc de véhicules (VANET). Ce travail présente nos premières investigations en décrivant la conception d'un système générique global. Le système conçu a été expérimenté à travers deux scénarios de services routiers basés sur la vidéo. En outre, l'architecture associée a été mise en \oe{}uvre sur un simulateur à échelle réduite d'un système embarqué véhiculaire. L'architecture développée a été testée dans le cas d'une application routière simulée visant à aider certains services de police. Le but de cette application est de reconnaître et de pister des véhicules et des individus recherchés en temps réel moyennant un système de surveillance formé par des véhicules en circulation. Le travail présenté démontre le potentiel de notre système pour améliorer efficacement et pour diversifier les applications nécessitant des traitements de vidéos en temps réel dans des environnements routiers.}\\
\\
{\textnormal{\textbf{Abstract}}}\\
{\small In this paper, we propose a design for novel and experimental cloud computing systems. The proposed system aims at enhancing computational, communicational and annalistic capabilities of road navigation services by merging several independent technologies, namely vision-based embedded navigation systems, prominent Cloud Computing Systems (CCSs) and Vehicular Ad-hoc NETwork (VANET). This work presents our initial investigations by describing the design of a global generic system. The designed system has been experimented with various scenarios of video-based road services. Moreover, the associated architecture has been implemented on a small-scale simulator of an in-vehicle embedded system. The implemented architecture has been experimented in the case of a simulated road service to aid the police agency. The goal of this service is to recognize and track searched individuals and vehicles in a real-time monitoring system remotely connected to moving cars. The presented work demonstrates the potential of our system for efficiently enhancing and diversifying real-time video services in road environments.}
\end{abstract}



\section{Introduction and Motivation}

In this work, we propose to exploit cloud computing systems for developing real-time road video services from embedded navigation systems and VANETs (Vehicular Ad-hoc NETworks). The proposed systems will have a final objective to be experimented on a vehicle fleet. More particularly, this paper presents the design, the implementation and the simulation parts of a cloud-based recognition system for extending real-time road video services. Indeed, the proposed global generic system will exploit a cloud-based embedded recognition systems and VANET technologies; on the one hand, for analyzing the road traffic (e.g.; vehicular or navigation information) and on the other hand, for mutualizing the computational resources as well as for sharing relevant information visually extracted. Notably, the designed system will be useful for identifying dynamical Points Of Interest from embedded cameras (e.g., traffic-based POI) and then sharing the identified POIs to external stakeholders potentially interested (e.g., surrounding vehicles or road agencies).\\

For instance, these technologies can be exploited for improving the road traffic, the emergency mapping or the citizen security by cooperatively analyzing acquired georeferenced road images. Respectively, we present below some scenarios that will be based on the detection of dynamical POIs:

\begin{itemize}
\item	\textbf{Sc.$1$:} a vehicle can detect an available parking area and to transmit its GPS location in a pre-defined neighborhood for informing surrounding drivers by exploiting a cloud computing system and VANET,
 
\item	\textbf{Sc.$2$:} each vehicle can similarly transmit images for analyzing and mapping the road meteorology in real-time. Thus, drivers can define an itinerary based on meteorological criteria, notably to reduce the moving in areas having bad weather (e.g., snowy roads),

\item	\textbf{Sc.$3$:} a vehicle can extract on-the-fly the license plate of preceding vehicles and then, sending the extracted plate characters to police services searching to localize stolen vehicles by matching extracted data with their reference databases,
 
\item	\textbf{Sc.$4$:} similarly, a vehicle can extract on-the-fly people faces from the streets and then, sending the extracted  face images to police services that aim to localize searched individuals.
 
\end{itemize}

In this study, we have experimented the proposed cloud-based system by considering the last scenarios related to the police service application.

\section{Related Work}

\begin{figure}[!t]
   \includegraphics[scale=0.5]{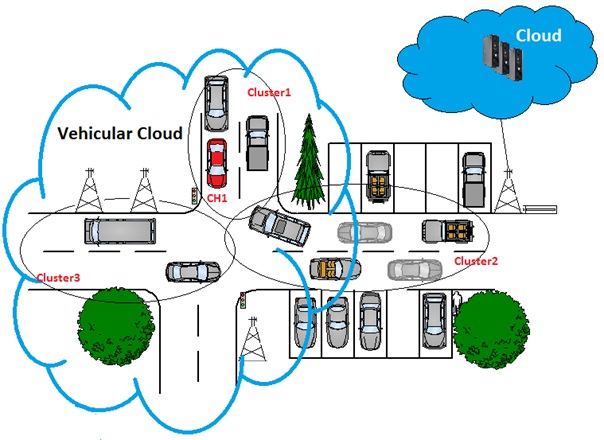}
	  \caption{Illustration of the designed cloud computing system.}
\label{fig:network1}
\end{figure}

Nowadays, cloud computing developments are revolutionizing the world by providing to companies more and more powerful services. In particular, many companies tend to store their data on external servers or data centers. Indeed, this technology improves the Quality of Service (QoS); notably for the data management, the data security as well as for the data distribution. By this way, the providers of cloud computing systems allow many companies to develop services specifically focused on their principal activities. More precisely, cloud computing can be defined as a technology providing resources at three levels, namely infrastructures, software platforms and services \cite{Whaiduzzaman}. The cloud computing was initially employed through wire-based network for internet and it has been progressively extended to the mobile network (e.g., through cellular networks). Notably, the cloud computing technologies facilitate the development of hybrid systems as well as the mutualizing of computational resources.

In this work, we are particularly interested by the development of cloud computing systems on the basis of VANET for enhancing and diversifying real-time road services. VANET networks have the particularity to exploit Ad-hoc systems. In other terms, these systems are self-organizing in the sense that each of them can communicate with others without the necessity of exploiting a pre-defined infrastructure. The development of VANET had a primary goal of supporting Intelligent Transport System through Vehicle-to-Infrastructure (V2I) and Vehicle-to-Vehicle (V2V) communications (e.g., \cite{Maslekar}). 

%

Besides, the novel generation of general public vehicles is equipped with computer-aided embedded navigation and vision systems such as Advanced Driver Assistance Systems (ADAS systems). In particular, ADAS systems are more and more employed for detecting road obstacles (e.g.; self-parking) or for detecting the visibility degree of roads (e.g.; automatic lighting systems). In parallel, experimental multi-camera vehicle systems are actively developed for the research in the fields of cartography and machine vision in order to reconstruct urban environments in 3D as well as to develop full autonomous navigation vehicles \cite{Hammoudi1, Hammoudi2}. 

%

To the best of our knowledge, video services in vehicular clouds are not very developed. In \cite{Gerla2013}, Gerla et al. presented an image-on-demand service named ``Pics-on-wheels'' where some vehicles will send their acquired images for example by analyzing detected accidents. These images can then be used for assurance claims. In our case, we present a generic cloud computing system that could be used for developing various real-time video services by exploiting a distributed computing system. Notably, this system will be employed for sharing traffic information (e.g.; in aided-navigation or road safety) by exploiting embedded vision-based systems (e.g., recognition system), CCSs and VANETs (see Figure \ref{fig:network1}).

\section{Proposed Global Generic System for Real-time Road Video Scenarios}

\begin{figure}[!h]
   \includegraphics[width=7cm]{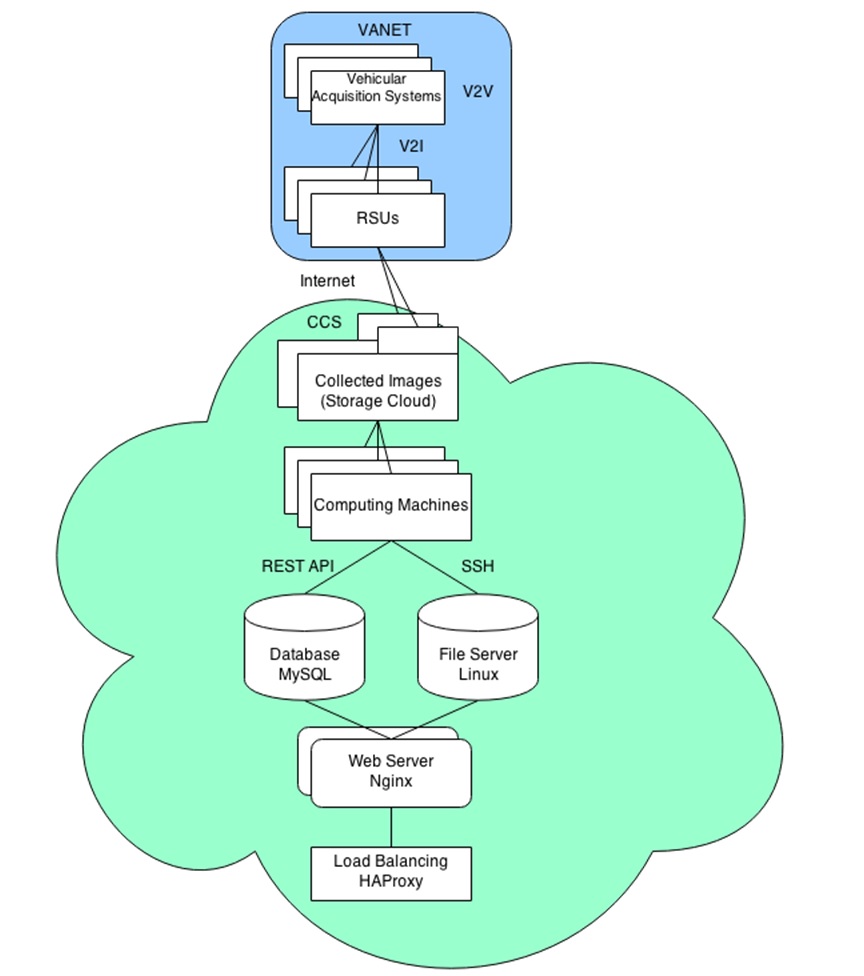}
	  \caption{Proposed processing architecture of a global generic system.}
\label{fig:diag2}
\end{figure}	

In our case, it is assumed that the vehicles will be equipped with embedded camera system, a GPS module and a VANET connecting system ($802.11p$). Notably, new generation vehicles are equipped with various types of sensors such as cameras located at the front and rear end. The proposed vision-based cloud computing system will take advantage of distributed computing and storing capabilities of conventional CCS and VANET (see Figure \ref{fig:network1}) for providing video services requiring high resources in term of data processing. In particular, the proposed system will be useful for visually recognizing dynamical objects of interest such as, for the search of stolen vehicles or individuals.

More precisely, the proposed system will exploit vehicular networks or external data center according to the needs. Yu et al. classify some cloud-based systems related to VANET \cite{Yu2013}. First, vehicular cloud is exclusively composed of vehicles. It allows vehicles to dynamically schedule on demand computational and storage resources. Second, roadside cloud is composed of dedicated servers and RSUs (Road Side Units). The later permits access to the cloud. This cloud is exclusively used by vehicles localized within the radio coverage of the RSU. Vehicles roam between successive RSUs to continuously benefit from the service. Third, central cloud is based either on dedicated servers in the Internet or data centers on VANET itself. In our case, we are using the concept of Hybrid Vehicular Cloud (HVC) which shares the processing between the Vehicular Cloud (VC) and the central cloud. 

Moreover, we visualize in Figure \ref{fig:diag2} the architecture that has been developed for supporting the various data transfer and data processing. First, vehicles communicate with internet access point by using vehicle to infrastructure (V2I) or vehicle to vehicle (V2V) communications. RSUs are exploited for removing redundancy in captured images and GPS information. 
Second, the collected georeferenced raw data are then sent to a customized storage cloud (e.g.; Amazon cloud). Computing machines continuously run the face extraction, GPS extraction and number plate recognition algorithms in parallel. The extracted license plate numbers as well as the extracted GPS information are saved in a database (textual information). The extracted images are copied to file servers. Users access the service by connecting to a load balancing server, which distributes the requests to several working web servers. 

\begin{figure}[!t]
   \includegraphics[width=7cm]{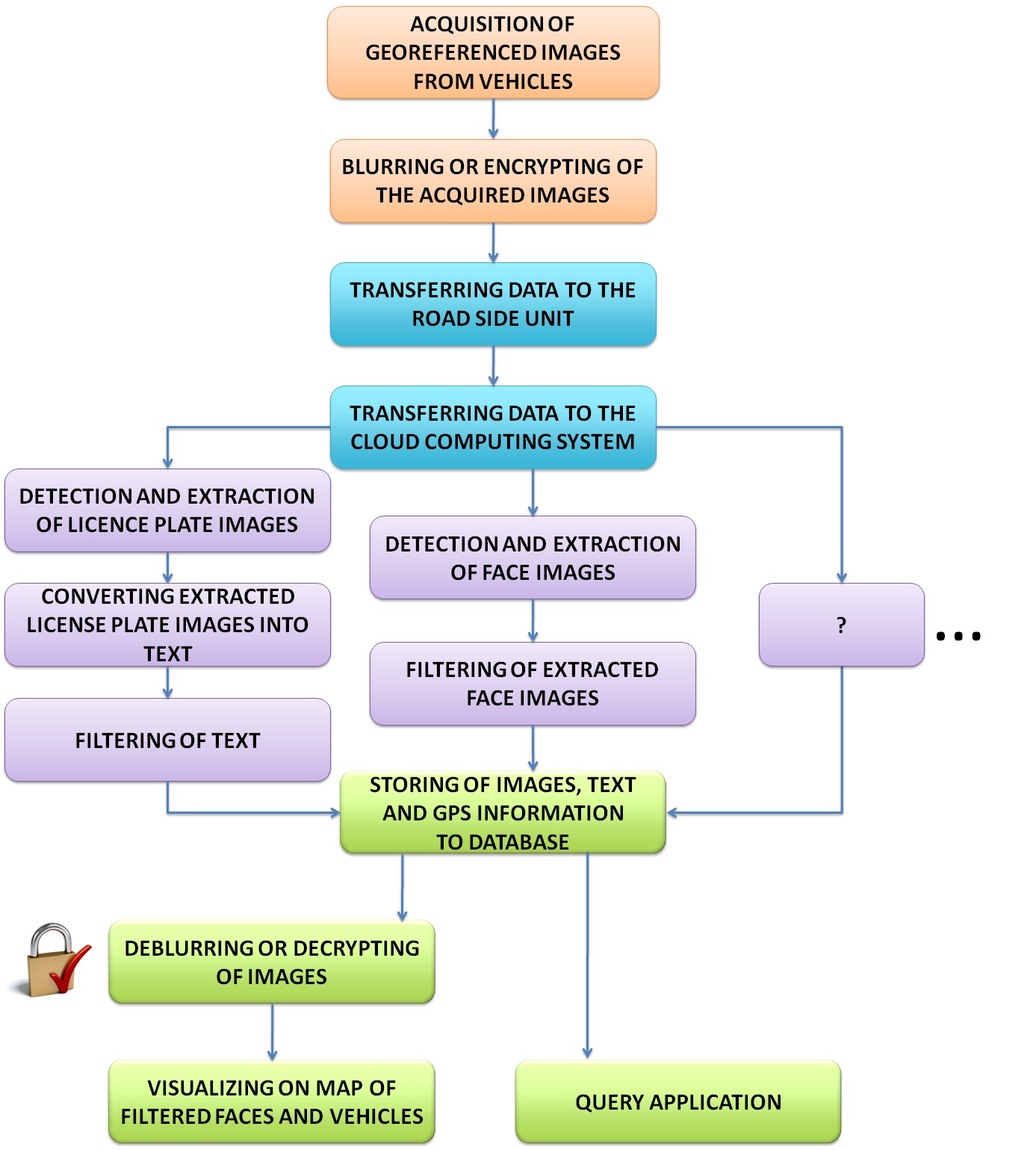}
	  \caption{Proposed global dataflow diagram.}
\label{fig:diag1}
\end{figure}	

In Figure \ref{fig:diag1}, we observe the global dataflow diagram of experimented scenarios (Sc.$3-4$).
As can be observed, it worth mentioning that our architecture can also be used for the processing of other scenarios related to new real-time road video services (e.g.; Sc.$1-2$).

\section{Experimental Results}

\subsection{Developed Indoor Vehicular Monitoring Simulator}

%
%

\begin{figure}
        \centering
        \begin{subfigure}[b]{0.40\textwidth}\centering
                \fbox{\includegraphics[scale=0.35]{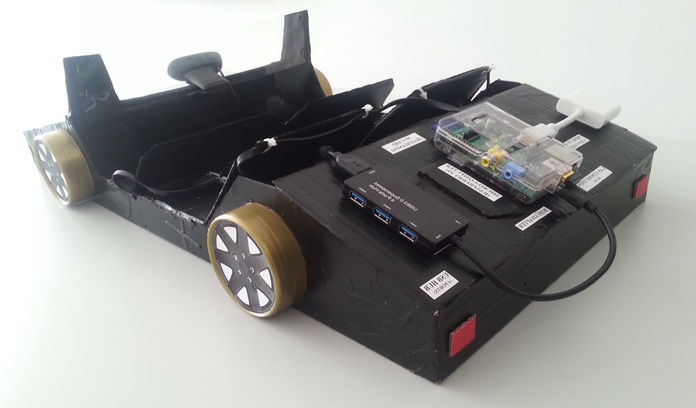}}
                \caption{Global view of the car-based monitoring simulator.}
                \label{fig:maquette}
        \end{subfigure}\\%
        ~ 
        \begin{subfigure}[b]{0.20\textwidth}\centering
                 \fbox{\includegraphics[height=2cm]{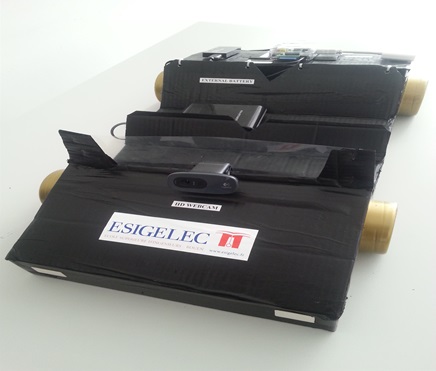}}
                \caption{Acquisition part.}
                \label{fig:simulator_front}
        \end{subfigure}
        ~ 
        \begin{subfigure}[b]{0.20\textwidth}\centering
                 \fbox{\includegraphics[height=2cm]{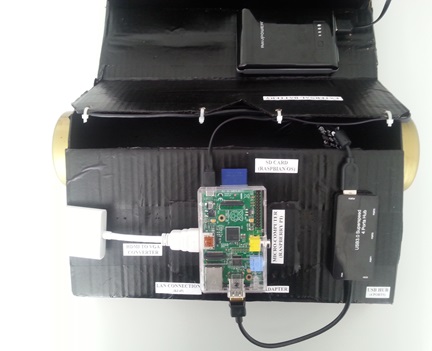}}
                \caption{Processing part.}
                \label{fig:simulator_top}
        \end{subfigure}
        \caption{Illustration of the developed car prototype and its associated vision-based embedded system (indoor vehicular monitoring simulator).}\label{fig:simulator}
\end{figure}

Figure \ref{fig:simulator} illustrated the developed embedded vehicular monitoring simulator. This vehicular monitoring simulator is composed of a car prototype (a rigid mock-up) as well as its associated vision-based embedded system (see Sub-figure \ref{fig:maquette}). This embedded system is equipped with a Logitech HD camera (see Sub-figure \ref{fig:simulator_front}) connected to a Raspberry Pi micro-computer (see Sub-figure \ref{fig:simulator_top}). This micro-computer includes a SD card for storing the acquired images. For simulating the moving of the car prototype, a screen has been placed in front of the webcam and a video corresponding to a vehicle path acquired by an external Mobile Mapping System has been filmed (e.g., videos from the Kitty research dataset\footnote{\url{http://www.cvlibs.net/datasets/kitti/}} \cite{Geiger2012CVPR,Fritsch2013ITSC}). For such databases, the GPS information related to the images are provided. The micro-computer includes a wifi adapter that was used for simulating the VANET network. This embedded car prototype is connected to three workstations, the one simulating the RSU, the two others simulating the cloud nodes.

\subsection{Implemented Architecture of the Proposed Global Generic System}

More precisely, two python scripts are running on Raspberry Pi, one aims to capture images and to geo-tag them, and the other aims to transfer the images to RSU by FTP. In RSU, a bash script is written to send those images to two simulated cloud nodes by using SSH. By this way, the data flow is evenly distributed to the cloud nodes through WiFi. The computing machines (also cloud nodes) will process the images in storage servers and get the extracted faces, license plates and GPS information by running a python script invoking the corresponding algorithms. On the web server, we implemented a RESTful API to access the database. The extracted images are archived in file servers, while the license number, GPS and time are updated to the database. Thus, the updated information can be visualized. Moreover, new extraction algorithms can be developed for various query applications.  

\subsection{System Application and Evaluation}

In this study, applications related to police services previously mentioned (Scenarios $3-4$) have been experimented by deploying computer vision approaches well-known for their efficiency on the proposed generic processing architecture (one simulated mobile node). Notably, open-source CSharp Emgu CV routines\footnote{\url{http://www.emgu.com/}} have been exploited for carrying out the face extraction as well as the OCR-based license plate extraction. Data matching has been experimented by comparing extracted features with a reference database generated by an operator. The proposed experimentation pipeline distributes the flow of collected images and extracted features are localized and labeled on Google Maps-based application in quasi real-time. 
Time information associated to the data transfer and data processing for one image of $16.5Kb$ (resolution of $640$x$480$) can be observed in Table \ref{tab1}.
 
\begin{table}
\centering \tiny
\begin{tabular}{|c|c||c|c|}
  \hline
  \textbf{Image transferring} & \textbf{Time (sec.)} & \textbf{Image processing} & \textbf{Time (sec.)} \\  
	\hline
	\hline
  Rasberry Pi to RSU & 1.33 & Face extraction & 1.08 \\
  RSU to cloud nodes & 1.12 & License plate extraction & 3.29 \\
	\hline
\end{tabular}
\caption{Time information associated to the data transfer and data processing for one image. Data is processed on Intel Core i5 workstations of $2.4GHz$ under Windows $8.1$ $64-bit$ with $4GB$ of RAM.} \label{tab1}
\end{table}

\section{Conclusions and Future Works}

\begin{figure}[!h]
   \includegraphics[scale=0.5]{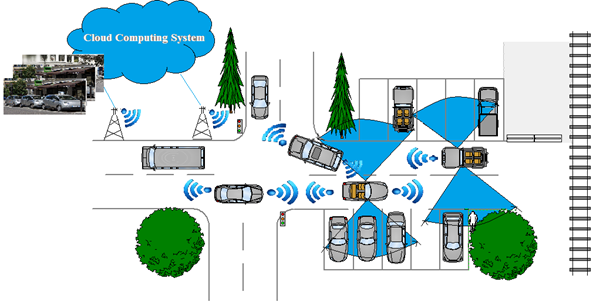}
	  \caption{Illustration of a scenario related to the detection of available parking areas by exploiting VANET.}
\label{fig:network2}
\end{figure}

This paper presents our initial investigations for the design, the implementation and the simulation of a cloud computing system for enhancing and diversifying real-time video services through VANET and Onboard Navigation Systems. A vehicular monitoring simulator has been developed for carrying out indoor experiments. A generic hardware and software architecture is proposed for experimenting new video service applications. 

Accordingly, next stage will consist of transferring this technology on two modular chassis that will be fixed on vehicle windshields for experiments in real mobile conditions (i.e., two moving nodes). Moreover, research will be pursued in indoor for improving the architecture of the developed simulator and simulations of the network architecture will be implemented under $ns2$ and $ns3$ Network Simulators\footnote{\url{http://nsnam.isi.edu/nsnam/}}\footnote{\url{http://www.nsnam.org/}}. Furthermore, we will tackle research in imagery for the detection of available parking areas in order to develop parking services. A corresponding targeted application was described in Scenario $1$ and has been illustrated in Figure \ref{fig:network2}. 

\section{Acknowledgements}

{\small This work is part of the SAVEMORE project\footnote{\url{http://www.savemore-project.eu/}}. The SAVEMORE project has been selected in the context of the INTERREG IVA France (Channel) - England European cross-border co-operation programme, which is co-financed by the ERDF.}


\bibliographystyle{refig-alpha}
{
\bibliography{bibsample}
}

\end{document}